\documentclass[doublecol]{epl2} 

\usepackage[utf8]{inputenc}
\usepackage{epsf,graphicx}
\usepackage{multirow}
\usepackage{amsmath,gensymb,amssymb}
\usepackage[bottom]{footmisc}
\usepackage{lipsum}
\usepackage{dcolumn}
\usepackage{bm}
\usepackage{siunitx}
\usepackage[bottom]{footmisc}
\usepackage{dcolumn}
\usepackage{flushend}
\usepackage{xcolor}
\usepackage{lipsum}

\usepackage{hyperref}
\hypersetup{
pagebackref=true,
colorlinks=true, 
citecolor=magenta,
breaklinks=true, 
urlcolor= blue, 
linkcolor= blue, 
bookmarksopen=true, 
}

\usepackage{color}
\definecolor{linkcolor}{rgb}{0,0,0.6}

\newcommand{\eff}{_{\mathrm{eff}}}

\DeclareMathOperator{\cn}{cn}
\DeclareMathOperator{\sech}{sech}
\DeclareMathOperator{\Tr}{\textbf{Tr}}
\DeclareMathOperator{\ellipE}{E}
\DeclareMathOperator{\ellipK}{K}

\title{Experimental observation of periodic Korteweg-de Vries\\ solitons along a torus of fluid}
\author{Filip Novkoski\inst{1}\thanks{\email{filip.novkoski@u-paris.fr}} \and Chi-Tuong Pham\inst{2}\thanks{\email{chi-tuong.pham@upsaclay.fr}} \and Eric Falcon\inst{1}\thanks{\email{eric.falcon@u-paris.fr} (corresponding author)}}
\shortauthor{Filip Novkoski, Chi-Tuong Pham and Eric Falcon}
\institute{
\inst{1}Universit\'e Paris Cit\'e, CNRS, MSC, UMR 7057, F-75013 Paris, France\\
\inst{2}Universit\'e Paris-Saclay, CNRS, LISN, UMR 9015, F-91405 Orsay, France
}

\abstract{We report on the experimental observation of solitons propagating along a torus of fluid. We show that such a periodic system leads to significant differences compared to the classical plane geometry. In particular, we highlight the observation of subsonic elevation solitons, and a nonlinear dependence of the soliton velocity on its amplitude. The soliton profile, velocity, collision, and dissipation are characterized using high resolution space-time measurements.  By imposing {\em periodic boundary conditions} onto Korteweg-de Vries (KdV) equation, we recover these observations. A nonlinear spectral analysis of solitons (periodic inverse scattering transform) is also implemented and experimentally validated in this periodic geometry. Our work thus reveals the importance of periodicity for studying solitons and could be applied to other fields involving periodic systems governed by a KdV equation.  }

\begin{document}

\maketitle

\section{Introduction}
Since their first observation on the surface of water~\cite{Russell}, solitons have been widely
studied in various domains (including acoustics~\cite{Hao2001}, plasmas~\cite{ZabuskyPRL1965},
carbon nanotubes~\cite{Astakhova2004}, Bose--Einstein condensates~\cite{Pitaevskii2003,Gammal2006},
or blood vessels of living organisms~\cite{Yomosa1987}). Korteweg and de Vries (KdV) first provided
an analytical description of solitons~\cite{Korteweg}, which can be observed as either waves of
elevation~\cite{HammackJFM74} or depression~\cite{Falcon2002} on the surface of a fluid. Although
KdV solitons have mainly been investigated experimentally in rectilinear
geometries~\cite{HammackJFM74,Falcon2002,RemoissenetBook,Dauxois2006,GrimshawBook}, examples in both
curved and periodic media remain elusive.

A stable torus of fluid is a good experimental system to study solitons in a curved and periodic
geometry. We manage to create such a stable torus of liquid by means of an original technique. We
have previously studied linear waves propagating along the inner and outer torus
borders~\cite{Novkoski2021}. Here, using this technique, we experimentally discover unreported
periodic KdV solitons along a stable torus of liquid whose properties are fully characterized
(profile, velocity, collision, and dissipation), and described with an experimentally validated
model taking into account both the curved and periodic conditions. Our work thus paves the way to
observe other nonlinear phenomena such as wave turbulence~\cite{FalconARFM2022,RicardEPL2021}, and
soliton gas~\cite{Zakharov1971,ELPRL2005,Costa2014,RedorPRL2019,Suret2020} in this specific
geometry. Note that KdV solitons can be reached experimentally in curved geometries without
periodicity ({\it e.g.,} along the border of a liquid
cylinder~\cite{BourdinPRL2010,PerrardPRE2015,LeDoudicJFM2021}), whereas trials have been attempted
for periodic conditions in plane geometry ({\it e.g.,} in an annular water
tank~\cite{Shi1998,Elizarova2012}), as well as for a curved and periodic system but only in a
nonstationary regime and by applying a strong constraint to the liquid
ring~\cite{PerrardEPL2012,Ludu2019,Vatistas2019}.

Theoretical works on solitons have yielded advanced mathematical techniques to study solutions to
various integrable nonlinear equations [{\it e.g.,} KdV, Nonlinear Schr\"odinger (NLS),
Kadomtsev-Petviashvili], in particular the inverse scattering transform
(IST)~\cite{Gardner1967,Ablowitz1981,Drazin1989,OsborneBook}. This nonlinear spectral analysis has
been applied to experimental NLS solitons~\cite{Suret2020,OsborneBook}, but remain scarce for KdV
ones~\cite{Christov2009,Osborne1980,RedorPRF2021}, and, so far, have not been applied to a periodic
experimental system, a more complex setting which has recently received numerical and theoretical
attention~\cite{Osborne1986,Osborne1994,OsborneBook,Christov2012}.

\begin{figure}[t!]
  \centering
  \includegraphics[width=1\columnwidth]{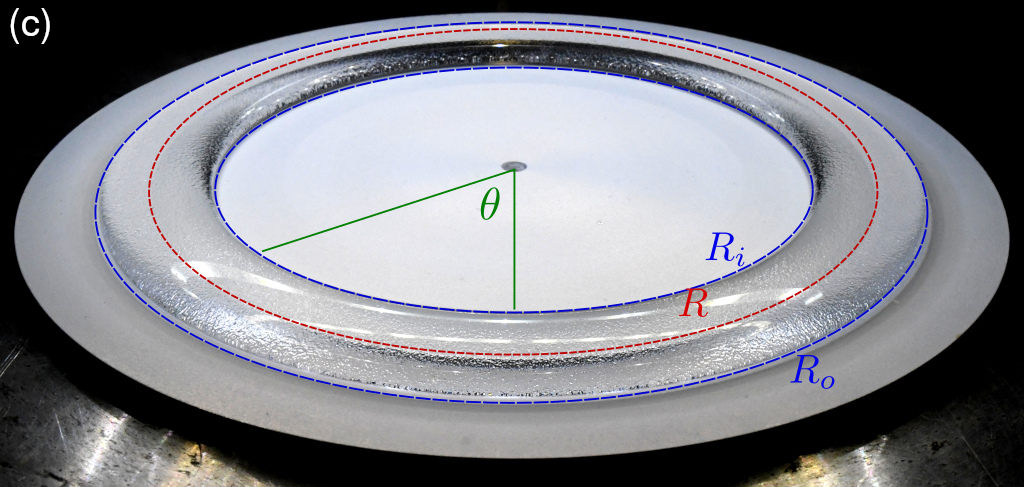} 
  \includegraphics[width=1\columnwidth]{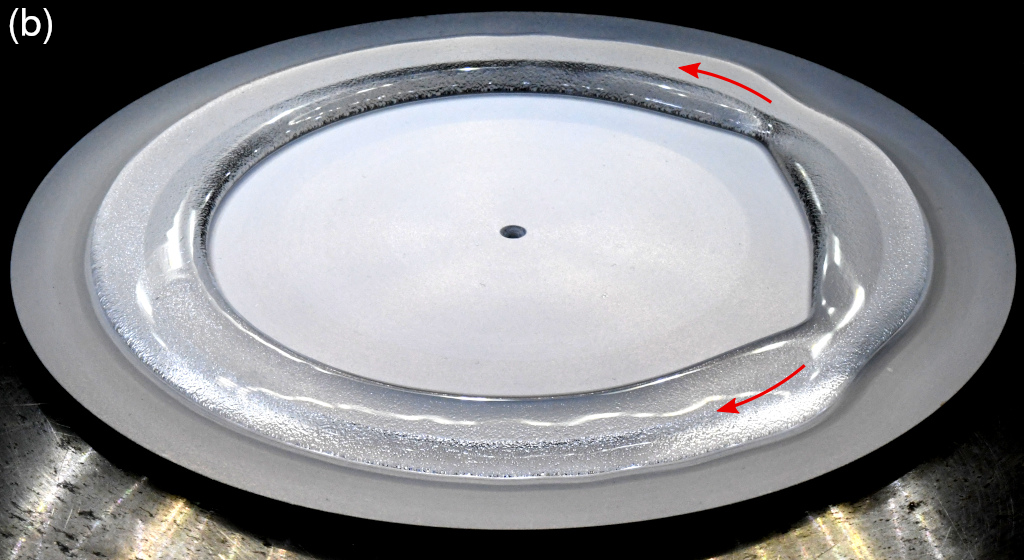} 
  \includegraphics[width=1\columnwidth]{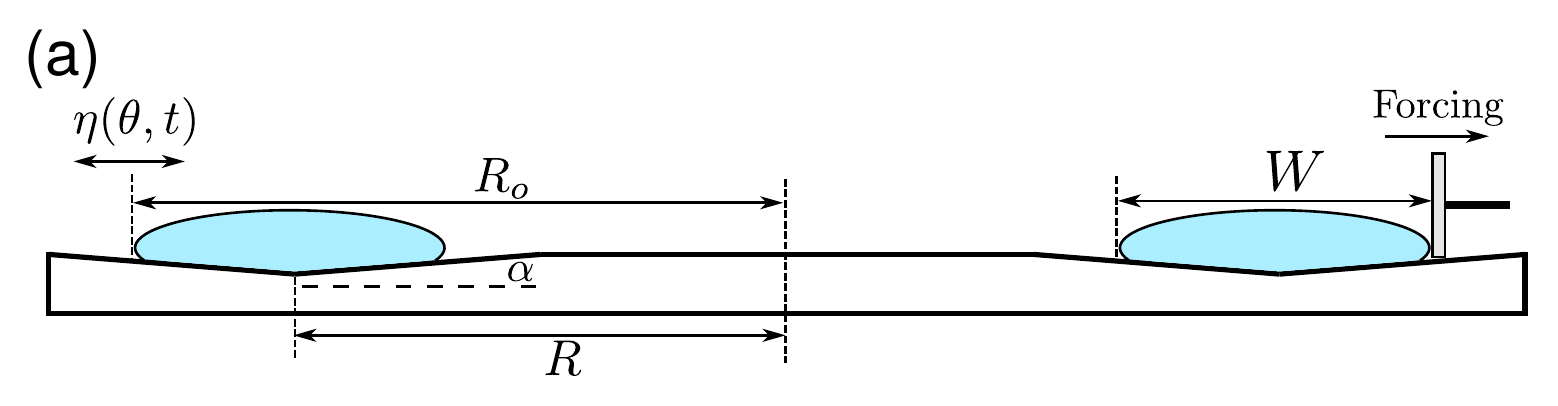} 
  \caption{a) Schematic profile of the experimental setup. b) Solitons propagating along the torus
    borders. c) Stable liquid torus on a plate ($R_o=7.9$\,cm, $R=7$\,cm, $W=1.8$\,cm).}
\label{fig:manip}
\end{figure}

\section{Experimental setup}
We manage to create a stable torus of fluid by depositing distilled water on a superhydrophobic
duralumin plate machined with a slightly sloping triangular groove along the perimeter (see
fig.~{\ref{fig:manip}a-c})~\cite{Novkoski2021}. The radius of the groove center, $R$, is either
$4$\,cm or $7$\,cm using two different substrates. The small angle $\alpha$ of the groove to the
horizontal is $4.5\degree$. We use a commercial superhydrophobic coating yielding a contact angle of
$160\degree$--$170\degree$ between liquid and substrate~\cite{Novkoski2021,Gupta2016} allowing the
liquid torus to move with almost no constraint. To generate waves, the torus is impulse pulled (or
pushed) horizontally using a linear actuator with a teflon plate attached to its end (see
fig.~\ref{fig:manip}a). By deforming the meniscus, the actuator creates two counter-propagating
solitons along the outer, and two along the inner, border of the torus (see fig.~\ref{fig:manip}b
and movies in Supp. Mat.~\cite{SuppMat}). A camera located above the torus records the displacements of the two interfaces. Using a border detection algorithm~\cite{github}, we extract the azimuthal
displacement $\eta(\theta,t)$, in the horizontal plane, of both the inner and outer torus
borders. 
Measurements are made for various pulse amplitudes and for different torus widths, $W$, by adding
water. We set $\chi=R_o/R$ in order to quantify the system curvature, with $R_o$ the outer radius of
the torus, and $R_o=R+W/2$ with $W$ the torus width (see fig.~\ref{fig:manip}a). KdV solitons in the toroidal geometry will be characterized using the $R=7$ cm case. The effects of periodicity on the solitons will be evidenced by decreasing the radius to $R=4$ cm.

\section{Soliton solutions}When weak dispersion is balanced by weak nonlinearity in a shallow
water regime, azimuthal waves $\eta(\theta,t)$ along a torus of fluid are governed at the leading order by
an ad hoc KdV equation with {\em periodic boundary conditions} as
\begin{align}\label{eq:kdv}
  \eta_t+\Omega_0\left[\eta_\theta+\frac{5\chi^2}{4\widetilde{W}}\eta\eta_\theta+\frac{\chi^2\widetilde{W}^2}{2R^2}\delta_{\mathrm{Bo}}\eta_{\theta\theta\theta}\right]=0\,{\rm ,}
\end{align}
with $\widetilde{W}=W/2$, $\delta_{\mathrm{Bo}}=\textup{Bo}_c-\textup{Bo}$, and
$\Omega_0={(g\eff \widetilde{W})^{1/2}/R}$ the angular phase velocity of linear gravity waves. The
Bond number reads $\textup{Bo}=\ell^2\eff /(\widetilde{W}^2\chi^4)$, $\textup{Bo}_c\approx1/6$, where $\ell\eff\equiv \sqrt{\sigma\eff/(\rho g\eff)}$ is the effective
capillary length, $\rho=10^3$\,kg\,m$^{-3}$ is the fluid density, $g\eff=g\sin\alpha$ is the
effective gravity, and $g=9.81$ m\,s$^{-2}$. $\sigma\eff\simeq 60$ mN\,m$^{-1}$ is an effective surface
tension inferred from the low-amplitude (linear regime) measurement of the dispersion relation. $g\eff$ and
$\sigma\eff$ are strongly linked to the substrate geometry and renormalization
effects~\cite{LeDoudicJFM2021}. We obtain eq.~\eqref{eq:kdv} using a Taylor expansion of the gravity-capillary dispersion
  relation along a liquid torus~\cite{Novkoski2021}, and adapting nonlinear corrections introduced in~\cite{LeDoudicJFM2021} for a rectilinear fluid cylinder to our torus geometry (see Supp. Mat~\cite{SuppMat}).

Cnoidal wave solutions to eq. \eqref{eq:kdv} read
\begin{align}\label{eq:profile}
  \eta(\theta,t)=A\cn^2\Bigg(\frac{\theta-\Omega t}{\Delta\sqrt{m}}\Big|m\Bigg)\ {\rm with}\ \Delta^2=\frac{24}{5} \frac{\widetilde{W}^3}{AR^2}\delta_{\mathrm{Bo}},
\end{align}
where $A$ is the (signed) amplitude and $\Delta$ the (angular) width of the solitary wave.  The sign of $A$ is given by that of $\delta_{\mathrm{Bo}}$.  The velocity of the soliton of eq.~\eqref{eq:profile} reads
\begin{align}\label{eq:velocity}
  \Omega=\Omega_0\left[1+\frac{5A}{6\widetilde{W} m}\chi^2 \left(1-\frac{m}{2}-\frac{3\ellipE(m)} {2\ellipK(m)} \right)\right]\,.
\end{align}
$\ellipK(m)$ [resp.  $\ellipE(m)$] is the complete elliptic integral of the first (resp.  second) kind. $m\in[0,1]$ is the elliptic parameter for which the cnoidal function $\cn(\theta |m)$ is $\cos(\theta)$ for $m=0$, and $\sech(\theta)$ for
$m=1$~\cite{GrimshawBook,Abramowitz1964}. Although the cnoidal wave is a periodic function, the $2\pi$-periodicity condition on the circle ({\it i.e.,} torus border) still has to be ensured, and reads
\begin{align}\label{eq:periodicity}
  \frac{2\pi}{N_\theta\Delta}=4K(m){\it ,\ i.e.,} \quad\pi = 2N_\theta\sqrt{\frac{6\tilde{W}^3m\delta_{\mathrm{Bo}}}{5R^2A}}\ellipK(m)\,,
\end{align}
with $N_\theta$ the number of solitons. The parameter $m$ and the amplitude  $A$ have thus a nontrivial relationship (see below).
The periodic elliptic solutions of eq.~\eqref{eq:profile} are close to $\sech^2$ for large enough
$R$ ({\it e.g.,} for $R=7$\,cm, $1-m\simeq10^{-12}$). 
In that case, eqs.~\eqref{eq:profile} and \eqref{eq:velocity} reduce to the classical solitary wave
profile $\eta(\theta,t) = A \sech^2[(\theta-\Omega
t)/\Delta]$ and velocity $\Omega=\Omega_0[1+5A\chi^2/(12\widetilde{W})]$.  However, for smaller
$R$ (e.g.  4\,cm), this classical solution cannot be used since the effect of periodicity, through
eq.~\eqref{eq:periodicity}, has to be taken into account (see below). Note that the
experimental parameters used here are in the range of validity required for the derivation of
eq.~\eqref{eq:kdv} assuming weak dispersion $\mu=\widetilde{W}^2\chi^2\delta_{\mathrm{Bo}}/(\Delta^2R^2)\in[0.05, 0.3]\ll1$ ({\it i.e.,} shallow-water limit), weak nonlinearity $\epsilon=A\chi^2/\widetilde{W}\in[0.005, 0.2]\ll 1$, both of the same order of magnitude $\mu/\epsilon=\widetilde{W}^3/(\Delta^2 R^2A)\in[1,3]$.

\begin{figure}[t!]
  \centering
  \includegraphics[width=1\columnwidth]{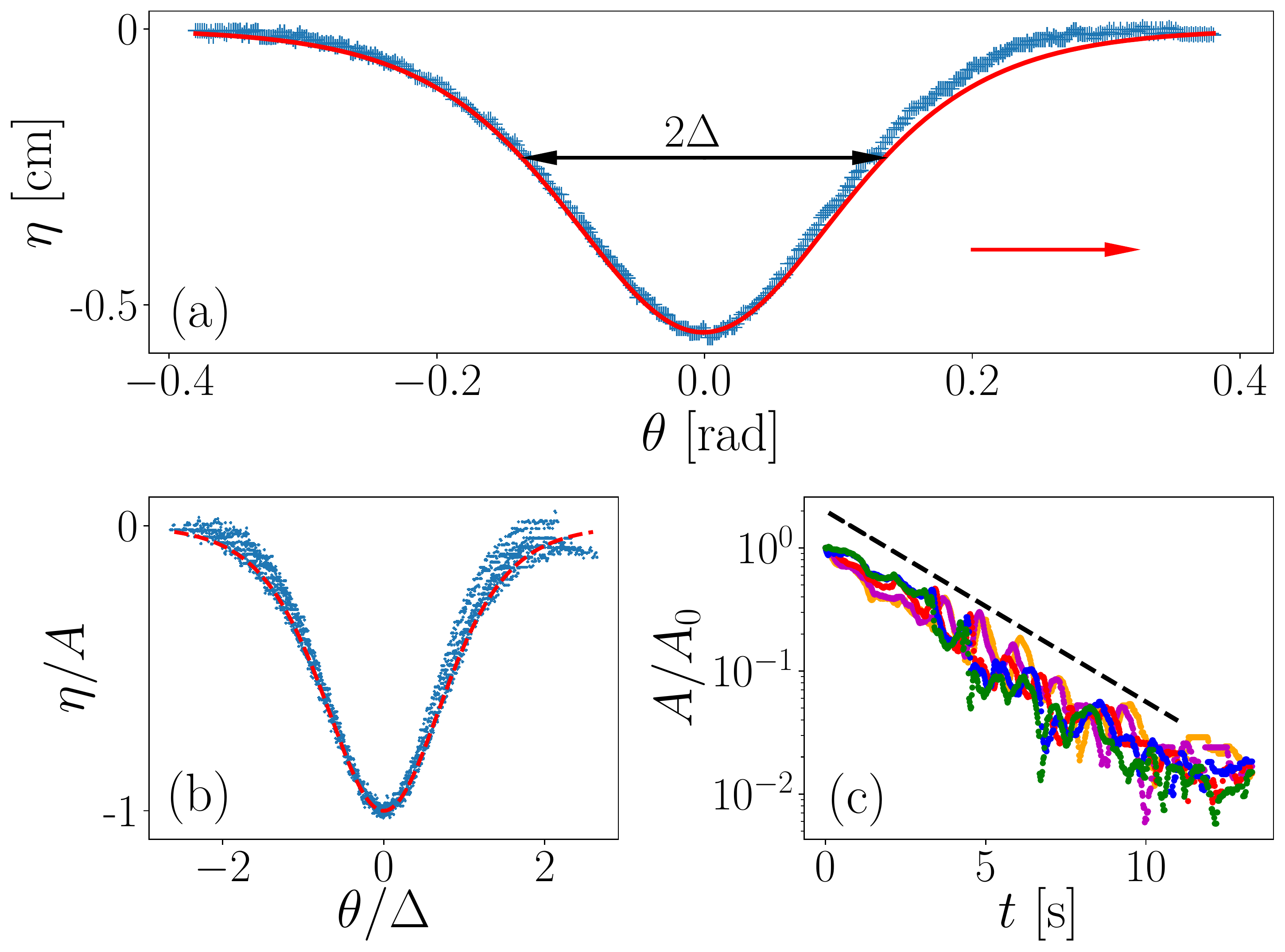} 
  \caption{a) ($+$) Experimental soliton profile at a fixed time. ($-$) Theoretical profile of
    eq.~\eqref{eq:profile} with no fitting parameter. b) Superimposition of rescaled soliton
    profiles during its propagation along one torus perimeter. ($-$) eq.~\eqref{eq:profile}. c)
    Exponential damping of the soliton for different $W\in[2.2,3]$\,cm (2\,mm
    step). $R=7$\,cm. Dashed line of slope $\tau=2.8$\,s.} 
\label{fig:profile}
\end{figure}

\section{Soliton profile}
The pulse profile, $\eta(\theta,t)$, is extracted from the outer torus border ({\it e.g.,} from the
depression in fig.\ \ref{fig:manip}b). Figure~\ref{fig:profile}a shows that the experimental profile
is well described by the theoretical soliton profile of eq.~\eqref{eq:profile} with no fitting
parameter. Since a soliton balances theoretically dispersion and nonlinearity, it should also have a
self-similar profile during its propagation. Figure~\ref{fig:profile}b shows the superimposed
rescaled profiles of a soliton during its propagation along almost one torus perimeter. The soliton
(with this appropriate rescaling) thus conserves a self-similar shape during its propagation that is
well described by eq.~\eqref{eq:profile}, even if its amplitude decreases due to unavoidable
dissipation. To quantify the latter, we plot in fig.~\ref{fig:profile}c the soliton amplitude as a
function of time, $A(t)$, during two rounds along the torus. $A(t)/A(0)$ is found to decrease
exponentially as $A(t)=A(0)\exp[-t/\tau]$, with a damping time $\tau$ found to be independent of the
viscosity of the fluid used ($\nu\in[10^{-7},10^{-6}]$\,m$^{2}$/s, {\it i.e.,} mercury or water).
This suggests that dissipation does not come from viscous dissipation, but probably from the pinning of the triple contact line \cite{Bonn2009}. Indeed, the capillary number ${\rm Ca=\rho\nu\Omega R/\sigma\eff}\in[10^{-6}, 10^{-3}]$ leads to dominant interfacial forces with respect to viscous ones.

\section{Fourier spectrum}
\begin{figure}[t!]
  \centering
  \includegraphics[width=1\columnwidth]{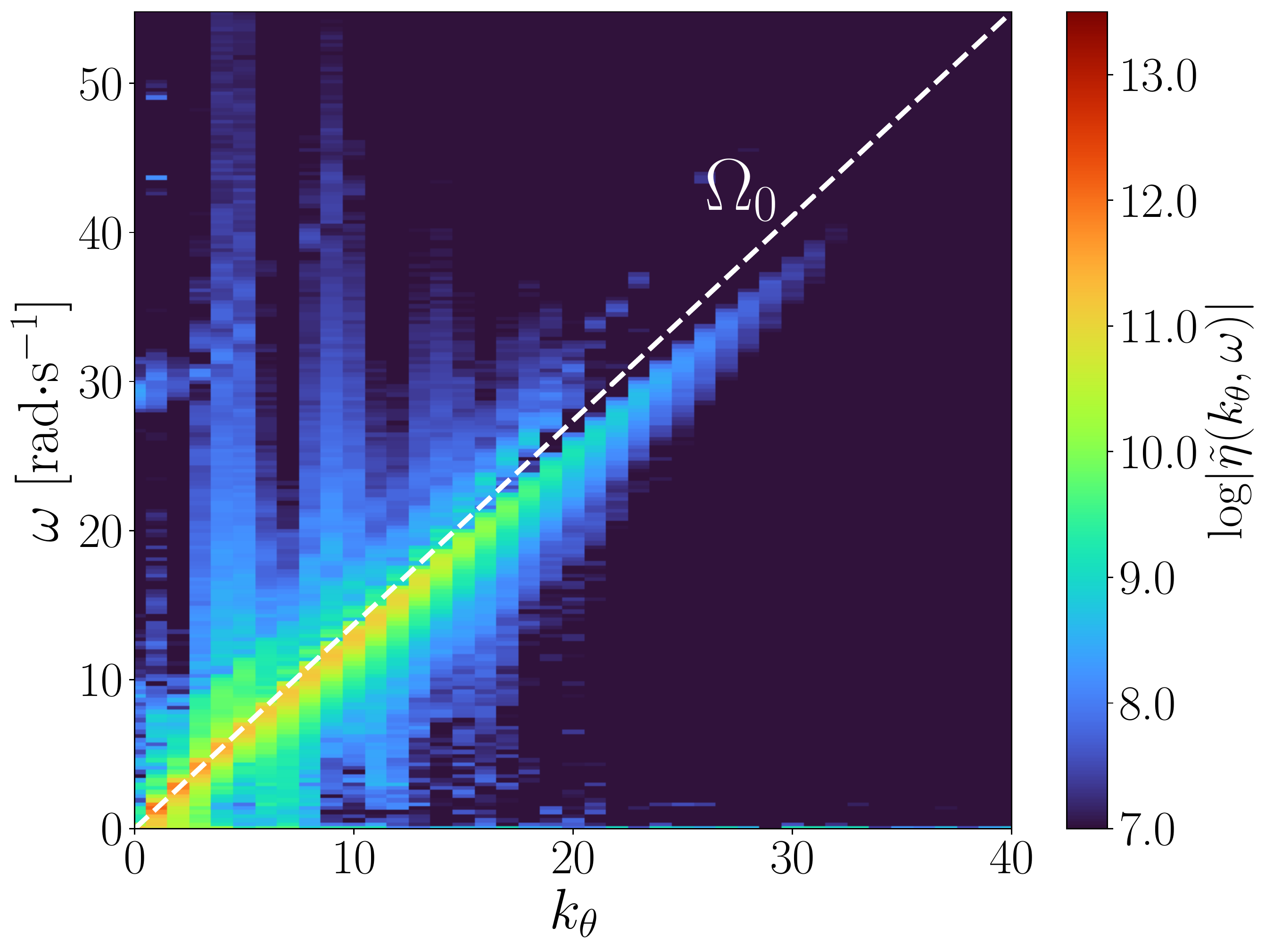} 
  \caption{Space-time Fourier spectrum $\tilde{\eta}(k_\theta,\omega)$ of the signal
    $\eta(\theta,t)$ (outer border). Dashed line: velocity $\Omega_0=1.37$ rad/s of long linear
    waves. The energy is concentrated around a linear branch of slope $\Omega<\Omega_0$, signature
    of a subsonic soliton.}
\label{fig:spectrum}
\end{figure}

We now compute the space-and-time Fourier transform, $\tilde{\eta}(k_\theta,\omega)$, of the signal
$\eta(\theta,t)$ as shown in fig.~\ref{fig:spectrum}. The energy is found to be concentrated around
a line of slope $\Omega=\omega/k_\theta$ corresponding to the pulse velocity. This
quasi-nondispersive feature is a spectral signature of a soliton.  The soliton velocity, $\Omega$,
is found to be slightly slower than long linear waves propagating at velocity $\Omega_0$ (see
fig.~\ref{fig:spectrum}), meaning the presence of a subsonic soliton. Note that a broadening of the
soliton branch occurs due to nonlinearities, whereas low-intensity vertical traces (at low
$k_\theta$) correspond to mechanical noise.


\begin{figure}[t!]
  \centering
  \includegraphics[width=1\columnwidth]{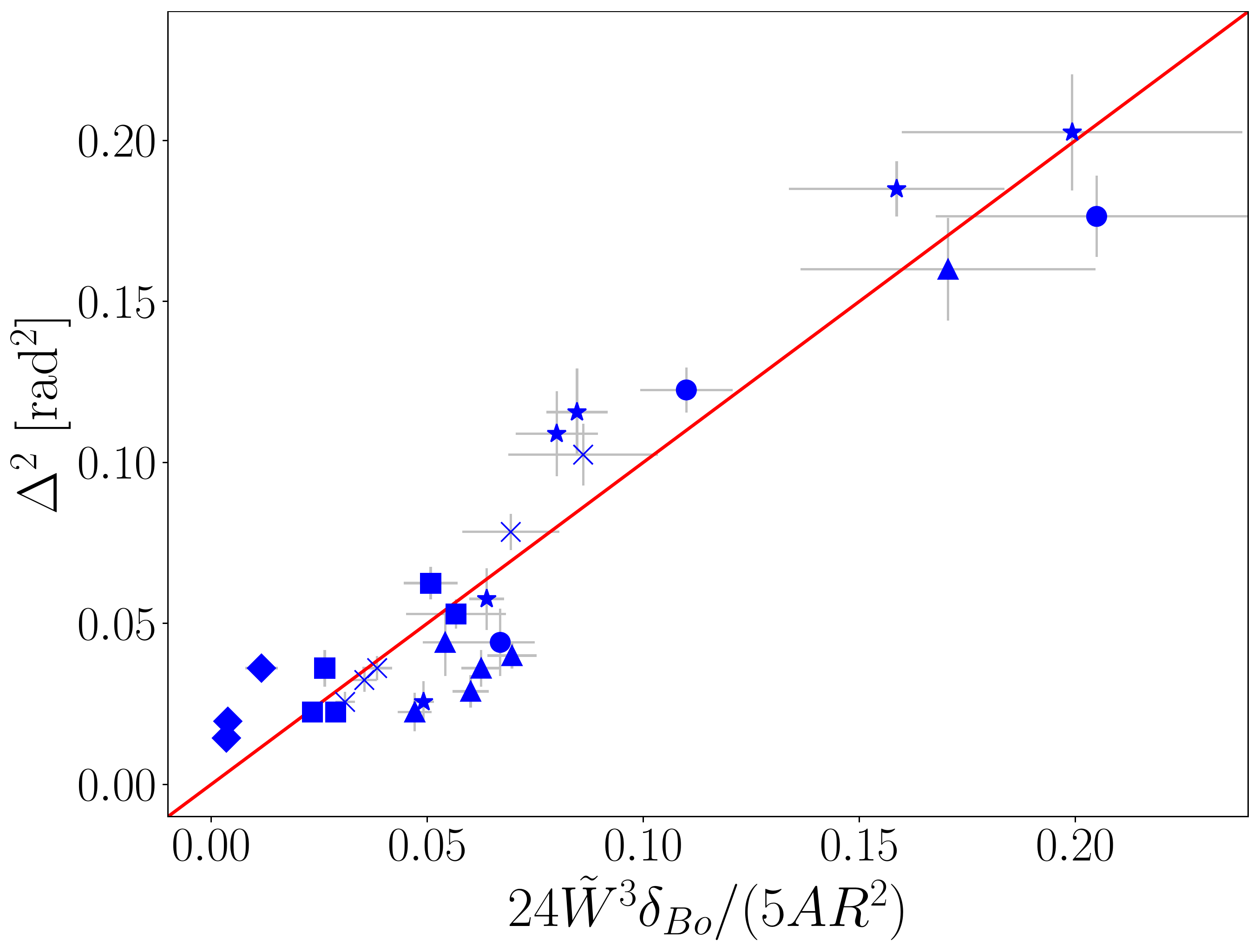} 
  \caption{Experimental soliton width squared $\Delta^2$ for different amplitudes $A$ and different
    widths $W\in[1.9,\,4]$\,cm (2\,mm step). $R=7$\,cm. Solid line: eq.~\eqref{eq:profile} with no
    fitting parameter (slope 1).}
\label{fig:delta}
\end{figure}

\section{Soliton width and velocity}
\begin{figure}[t!]
  \centering
  \includegraphics[width=0.9\columnwidth]{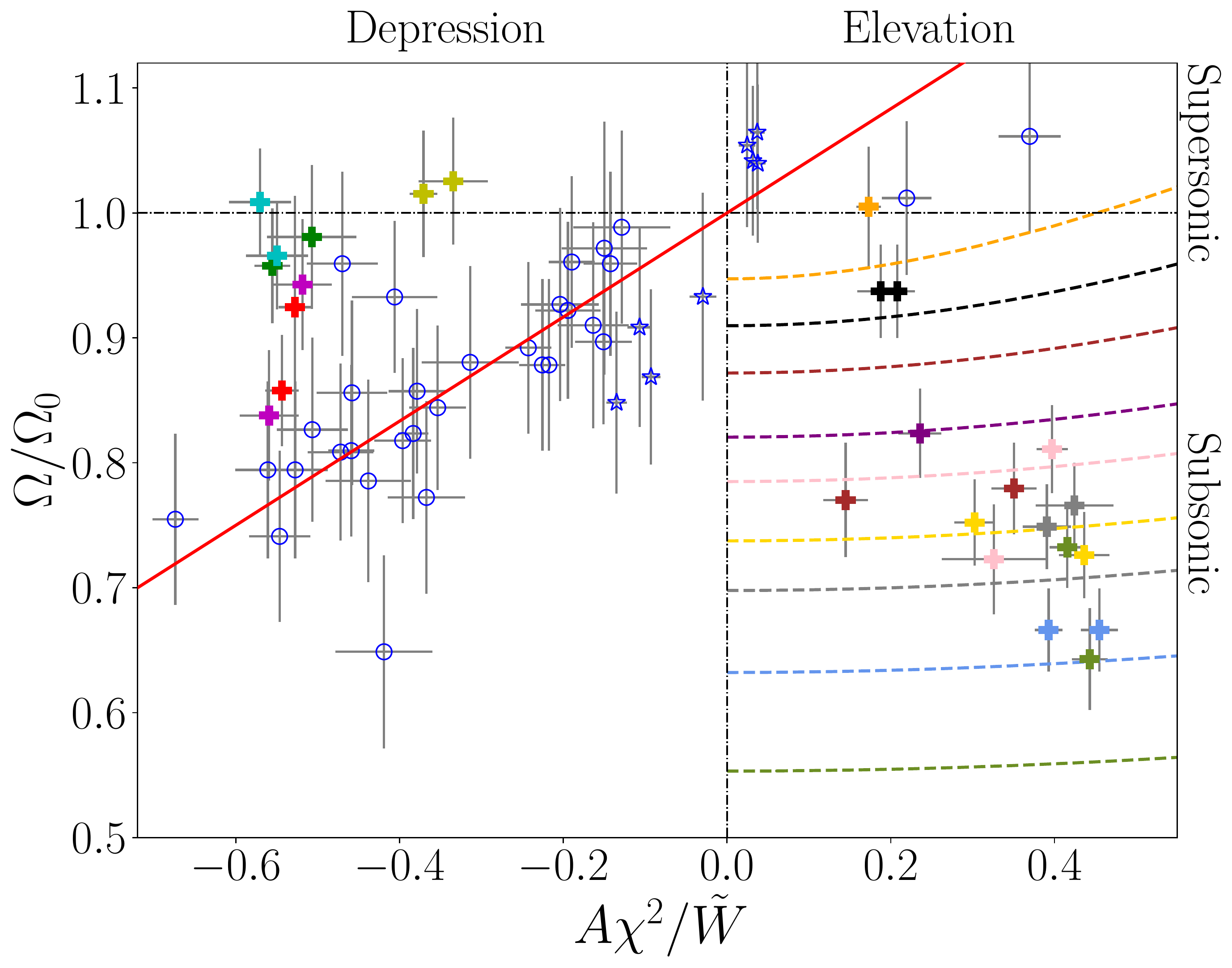} 
  \caption{Dimensionless soliton velocity $\Omega/\Omega_o$ versus $A\chi^2/W$ for various $A$ and
    $W$ for $R=$ ($\boldsymbol{+}$) 4 and ($\circ$) 7\,cm. Dashed lines: eq.~\eqref{eq:velocity} for
    different $ W\in[2.8,3.9]$. Solid line: classical KdV solution (slope 5/12). Occurrence of
    subsonic elevation solitons is due to effects of the periodic geometry.}
\label{fig:velocity}
\end{figure}

We now measure the typical soliton width $\Delta$ by fitting eq.~\eqref{eq:profile} to the
experimental profile (as in fig.~\ref{fig:profile}a). $\Delta^2$ is plotted in fig.~\ref{fig:delta}
for different pulse amplitudes, $A$, and torus widths $W$. $\Delta$ is found to scale as
$\sqrt{W^3/A}$ in good agreement with eq.~\eqref{eq:profile}b with no fitting parameter (see solid
line), thus justifying our ad hoc model that will lead to further predictions (see below). We
also measure the soliton velocity by time of flight during its propagation. The dimensionless pulse
velocity, $\Omega/\Omega_o$ ({\it i.e.,} Froude number), is displayed in fig.~\ref{fig:velocity} for
various $A$ and $W$. For large tori ({\it i.e.,} using the substrate $R=7$\,cm for various $W$), the
soliton velocity of eq.~\eqref{eq:velocity} reduces to the classical KdV linear
relationship, 
$\Omega/\Omega_0=1+5A\chi^2/(12\widetilde{W})$ (see solid line), which is well verified
experimentally (open circles). Depression solitons ($A<0$) moving slower than linear waves
($\Omega/\Omega_0<1$ or subsonic) are observed for $\textup{Bo} > \textup{Bo}_c$, whereas elevation
solitons ($A>0$) are supersonic ($\Omega/\Omega_0>1$) for $0\leq \textup{Bo} < \textup{Bo}_c$, as
predicted for KdV in straight geometry~\cite{Korteweg,Falcon2002}. For smaller tori ({\it i.e.,}
$R=4$\,cm substrate), the relationship of eq.~\eqref{eq:velocity} between velocity and amplitude is
no longer linear (see dashed lines from eq.~\eqref{eq:velocity} for different $W$). In particular, we clearly observe {\em subsonic elevation} solitons due to the effects of the periodic geometry (see $\boldsymbol{+}$ in the bottom right
quadrant).

\section{Periodicity effects on the soliton velocity}
\begin{figure}[t!]
  \centering
  \includegraphics[width=0.94\columnwidth]{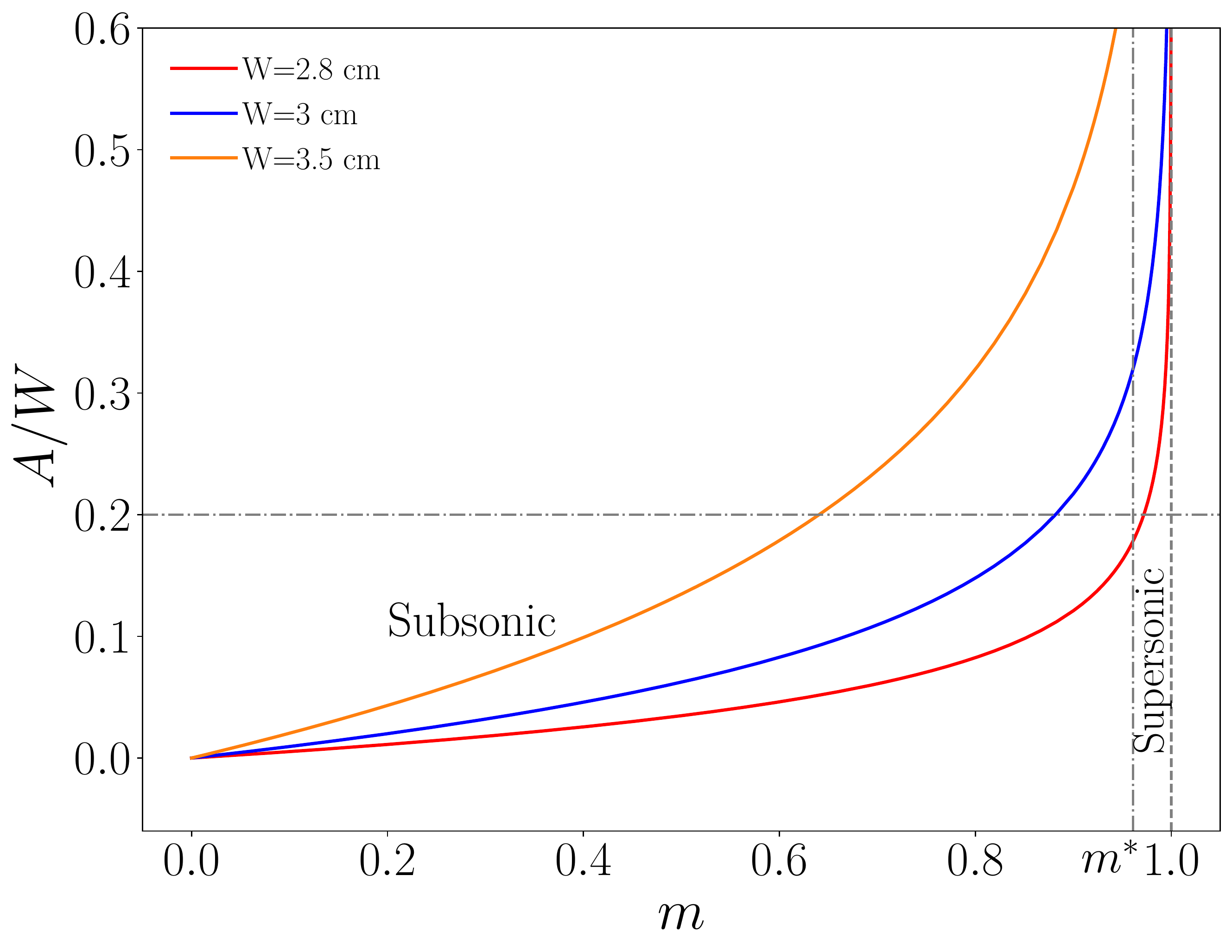} 
  \caption{Theoretical dependence of the elevation soliton amplitude $A$ on the parameter $m$ for different torus
      widths $W$, using eq.~\eqref{eq:periodicity} with $N_\theta=2$ and $R=4$\, cm. $m^*=0.96$ corresponds to the transition between subsonic and supersonic elevation solitons. The horizontal line ($A/W=0.2$) corresponds to the maximal soliton amplitude reachable experimentally. This thus limits attainable values of $m$, thus restricting the observation to the subsonic case for elevation solitons ($\delta_{\mathrm{Bo}}>0$, {\it i.e.,} $\mathrm{Bo} <\mathrm{Bo}_c$).}
\label{fig:am}
\end{figure}

The transition from subsonic to supersonic solitons occurs, from eq.~\eqref{eq:velocity}, at
$m^*=2-3E(m^*)/K(m^*)\simeq 0.96$ regardless of $\mathrm{Bo}$. This leads to different solutions of the
periodic KdV equation as follows
\begin{center}
\begin{tabular}{ l|l|l|l } 
  $\delta_{\mathrm{Bo}}$     &      Type            & $0<m<m^*$ & $m^*<m<1$ \\ \hline  
  $>0$ & Elevation  & Subsonic     & Supersonic \\ \hline  
  $<0$ & Depression & Supersonic & Subsonic   
\end{tabular}
\end{center}
An additional effect of the periodicity condition eq.~\eqref{eq:periodicity} is that certain types of solitons are unreachable experimentally due to our finite ranges of $W$ and of $A$. This can be seen by plotting the dependence of the soliton amplitude $A$ on the elliptic parameter $m$, $A/W=\frac{3}{5}mK^2(m) \delta_{\mathrm{Bo}}\left(\frac{N_\theta W}{\pi R}\right)^2$ from eq.~\eqref{eq:periodicity}, as shown in fig.~\ref{fig:am} for different widths $W$ for which elevation solitons are observed. Due to the finite size of the torus, the soliton amplitude is experimentally limited typically to $A/W<0.2$.
\begin{figure}[t!]
  \centering
  \includegraphics[width=1\columnwidth]{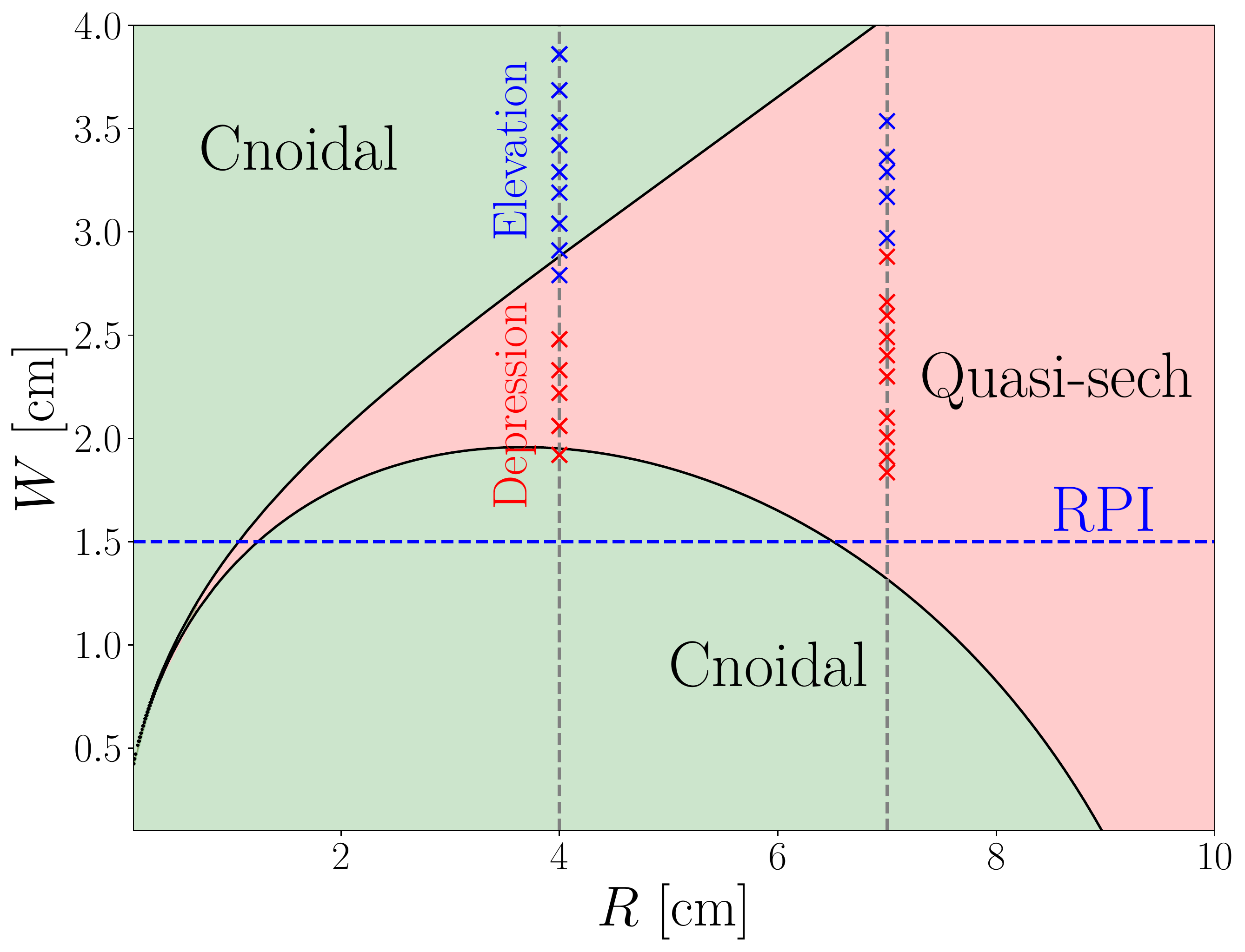} 
  \caption{Phase diagram of the different solutions of the periodic KdV equation. Solid lines: eq. \eqref{eq:periodicity} with $A/W=0.2$, $m=0.999$ and $N_\theta=2$. ($\times$) Torus widths for which depression (red) and elevation (blue) solitons are experimentally observed for the two tested substrates corresponding to a torus radius of $R=4$\,cm or $R=7$\,cm (vertical dashed lines). The torus minimal width is limited by the Plateau--Rayleigh instability (PRI).}
\label{fig:phase-diagram}
\end{figure}As $m$ increases with $A$, this also limits the reachable values of $m$, and thus the experimentally reachable solution types, as corroborated by the results of fig.~\ref{fig:velocity} ({\it e.g.,} no observation of elevation supersonic soliton for $R=4$\, cm). An equivalent plot to fig.~\ref{fig:am} can be obtained for the depression soliton case ($\delta_{\mathrm{Bo}}<0$, {\it i.e.,} $\mathrm{Bo} >\mathrm{Bo}_c$) provided that the subsonic and supersonic regions are swapped.

Figure \ref{fig:phase-diagram} sums up the experimentally observable cases in a phase diagram in the ($R$,$W$) parameter space. Inserting the experimental maximal soliton amplitude $A/W=0.2$, and $m=0.999$ [separating cnoidal soliton solutions ($m<0.999$) from quasi-sech ones ($m>0.999$)], into the periodicity condition of eq.~\eqref{eq:periodicity} leads to green regions for cnoidal soliton solutions and salmon-pink one for quasi-sech solutions. The experimental data for small tori ($R=4$\,cm) fall in both the cnoidal and quasi-sech regions whereas those for a large tori ($R=7$\,cm) fall completely in the quasi-sech region, justifying well the velocity observations in fig.~\ref{fig:velocity}. Note that, for $m$ far from $1$, we still refer to solutions as solitons since they experimentally propagate around the torus as solitary waves and undergo nonlinear interaction, although displaying a nontrivial amplitude dependence velocity (see fig.~\ref{fig:velocity}). It is worth noting that, according to eq.~\eqref{eq:periodicity}, the limit of $m=1$ is unreachable under periodic conditions since it would require an infinite amplitude.

\begin{figure}[t!]
  \centering
  \includegraphics[width=1\columnwidth]{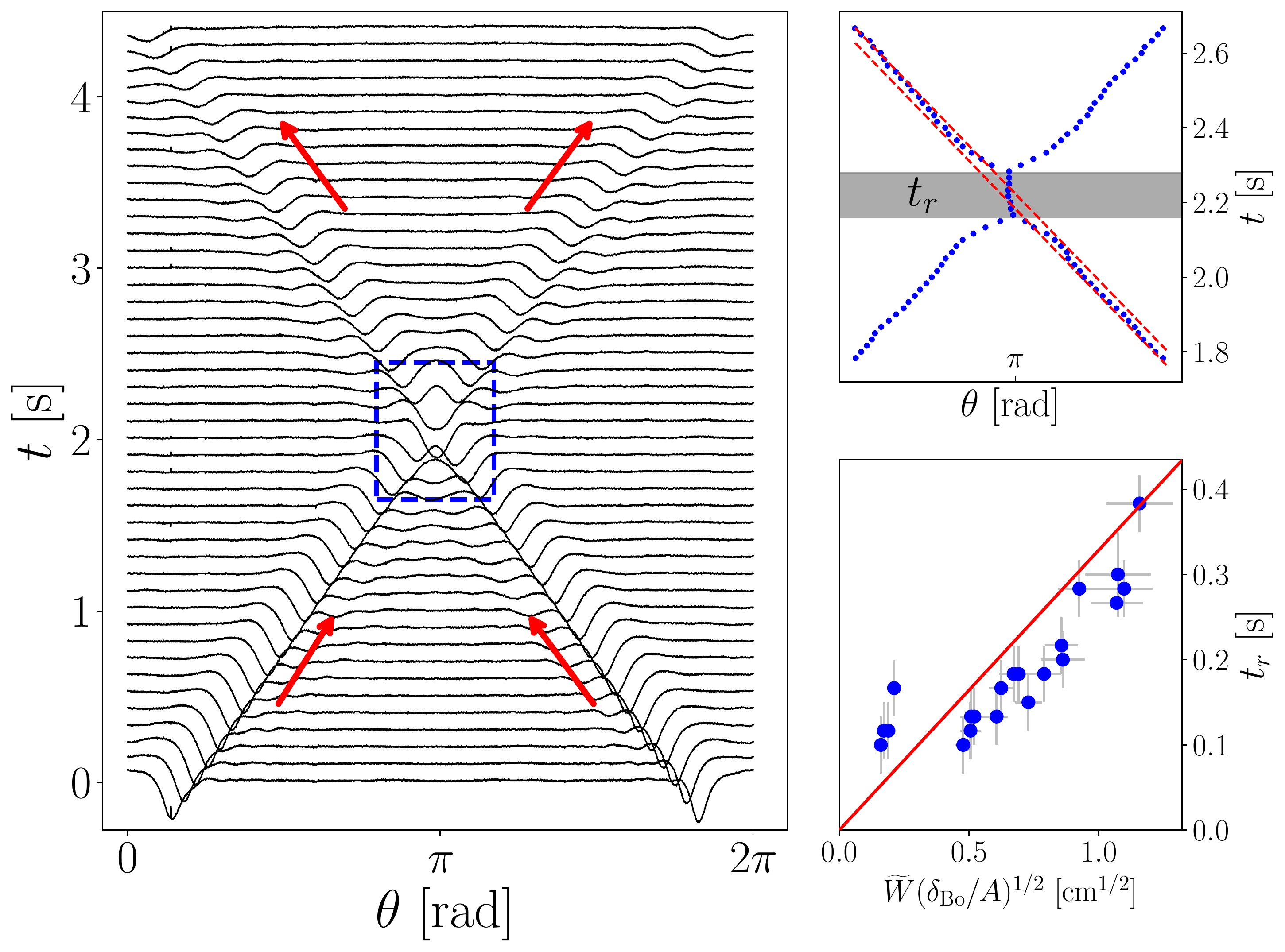} 
  \caption{Angle-time plot of a head-on collision between two depression solitons propagating
    along a torus. Top: Enlargement (dashed square) showing the soliton minima, 
    phase shift, and residence time $t_r$. Bottom: $t_r$ vs
    $\widetilde{W}\sqrt{\delta_{\textup{Bo}}/A}$ for various $A$ and $W$. Solid line slope is 0.33\,s/m$^{1/2}$. $R_o=8.2$\,cm.}
\label{fig:3d}
\end{figure}

\section{Critical Bond number}
The critical Bond number corresponds to the transition between elevation and depression soliton
solutions~\cite{Falcon2002}. It is remarkable that the theoretical value of the critical Bond number
$\textup{Bo}_c\approx 1/6$ for a torus (see Supp. Mat.~\cite{SuppMat}) differs from the value $1/3$ for the plane geometry
case~\cite{Korteweg}. Indeed, $\textup{Bo}_c$ strongly depends on the substrate slope $\alpha$ as found
numerically~\cite{LeDoudicJFM2021}. Equating the Bond expression to $1/6$ and inserting
$\widetilde{W}=R_o-R$, we find the critical outer radius $R^c_o$ of the torus separating elevation
and depression solitons as $R^c_o{^3}-R^c_o{^2}R-\sqrt{6}\ell\eff R^2=0$, and thus $R^c_o=8.43$\,cm
for our parameters. Experimentally, we have a range of $\mathrm{Bo}\in[0.09,0.5]$ by varying $R_o$,
and we look for the occurrence of the transition from depression ($R_o< R^c_o$) to elevation
($R_o> R^c_o$) solitons by increasing $R_o$. For small $R_o$, depression solitons are indeed
observed, whereas elevation solitons are detected above a certain radius. We find a critical
experimental radius of $R^c_o = 8.4\pm0.02$\,cm in good agreement with the above predictions. This
corresponds to $\textup{Bo}_c=0.17$ close to the theoretical value $1/6$. This result is also
confirmed when using the other substrate ($R=4$\,cm).

\begin{figure}[t!]
\centering
  \includegraphics[width=1\columnwidth]{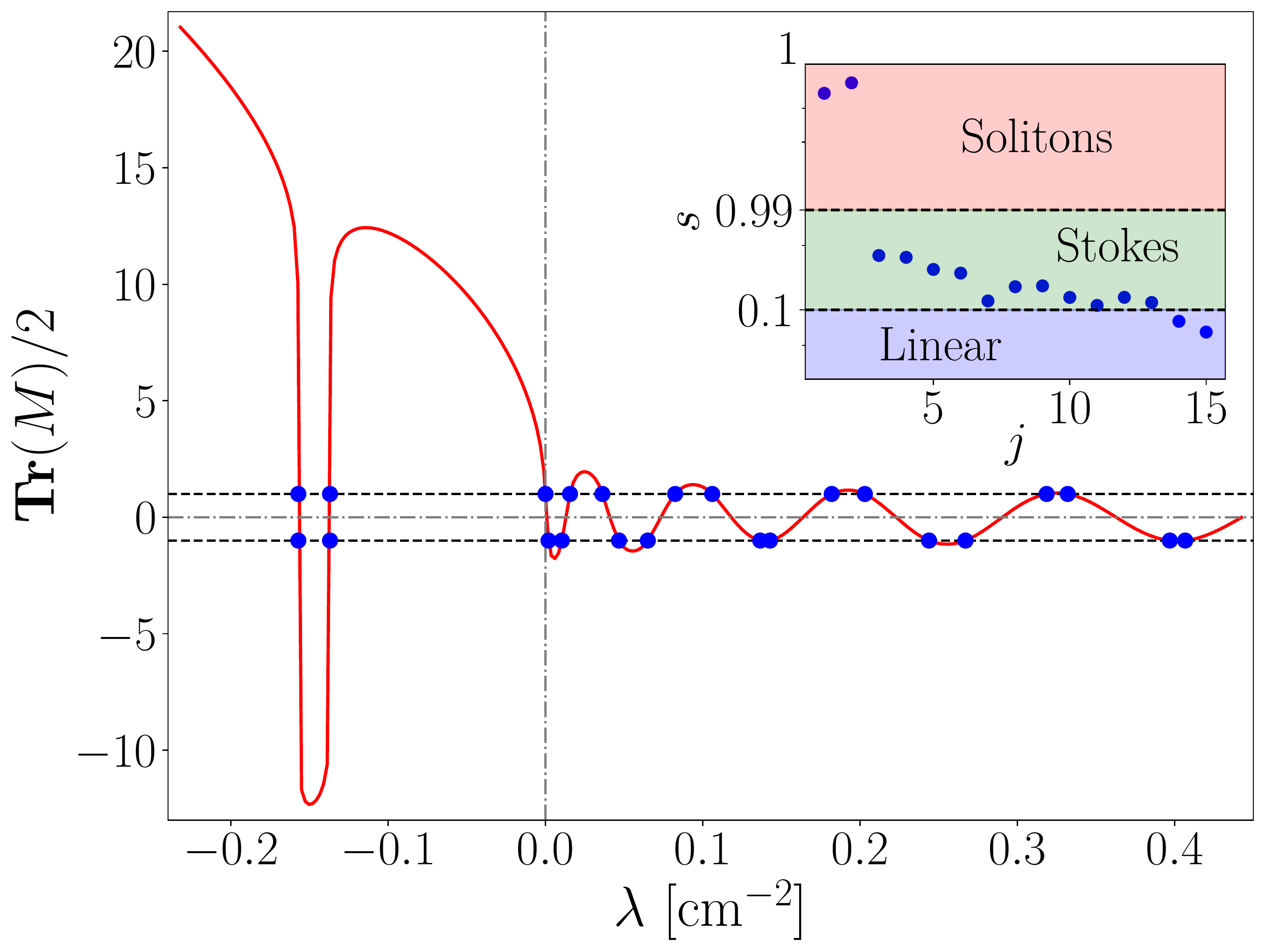} 
  \caption{PIST detection of solitons. $\Tr[M(\lambda)]/2$ (red line) for the signal of
    fig.~\ref{fig:3d} at $t_0=2.7$\,s, with the associated nonlinear spectrum
    (bullets). Two solitons are detected (bullets with $\lambda<0$). Inset: Soliton index $s$ for
    different azimuthal wave numbers $j$ revealing two solitons ($s>0.99$). Lin-Logit scale.}
  \label{fig:pist}
\end{figure}

\section{Soliton collision}
The nonlinear nature of the solitons is further confirmed by observing the collisions of two
depression solitary waves as illustrated in fig.~\ref{fig:3d}. Figure~\ref{fig:3d} (top right) shows an enlargement of
the two solitary wave minima as they collide. The collision evidences a long residence time $t_r$
(of the order of $0.1$\,s) during collision, and a slight phase shift, a feature of solitons.  We experimentally
show in fig.~\eqref{fig:3d} (bottom right) that $t_r$ scales as
$t_r\sim \widetilde{W}\sqrt{\delta_{\textup{Bo}}/A}$, matching our prediction (see
Supp. Mat.~\cite{SuppMat}) and extending the pure gravity prediction~\cite{Power1984}.

\section{Direct scattering}
We have shown above that solitons observed along a liquid torus are well described by
eqs.~\eqref{eq:profile}-\eqref{eq:periodicity}, the solutions of the periodic KdV eq.~\eqref{eq:kdv}. We
now implement a nonlinear spectral analysis, using the periodic inverse scattering transform (PIST),
to find the discrete eigenvalue $\lambda$ of each soliton in our signals~\cite{Drazin1989}. To the best of our knowledge, such a method has not been applied so far to an experimental periodic system with a significant
discreteness in Fourier space. We associate with eq.~\eqref{eq:kdv} the following eigenvalue
problem~\cite{Drazin1989,Osborne1986}
\begin{align}\label{eq:schrodinger}
  \psi_{xx}+\left[\beta\eta(x,t=t_0)+\lambda\right]\psi=0\,,
\end{align}
subjected to periodic boundary conditions, with period $L=2\pi R_o$, and
$\beta=5/(12\widetilde{W}^3\chi^2\delta_{\mathrm{Bo}})$. The eigenvalues correspond to either
bounded solutions, {\it i.e.,} solitons, for $\lambda<0$, and Stokes waves or radiative phonons for
$\lambda\geq 0$~\cite{Ablowitz1981}. We use a periodic scattering matrix $M(\lambda)$ (called
monodromy matrix) to translate the solutions of eq.~\eqref{eq:schrodinger} by one period. The
nonlinear spectrum is then given by the condition $\Tr[M(\lambda)]/2=\pm1$. The experimental
nonlinear spectrum is displayed in fig.~\ref{fig:pist} (bullets), along with the half-trace of the
matrix $M$ (solid line) for the signal in fig.~\ref{fig:3d} at a time $t_0$. Two solitons are
detected in fig. \ref{fig:pist} for which $\Tr[M(\lambda)]/2=\pm1$ (four eigenvalues or two
band gaps), corresponding to two distinct values $\lambda<0$. From this nonlinear spectrum, we
compute the soliton index $s$, for each nonlinear mode, as~\cite{Christov2009}
\begin{align}
s=\frac{\lambda_{2j+1}-\lambda_{2j}}{\lambda_{2j+1}-\lambda_{2j-1}}\,,  
\end{align}
which corresponds to solitons if $s>0.99$, Stokes waves if $0.5<s<0.99$, or linear radiative modes if
$s<0.5$~\cite{OsbornePRE1995}. We are thus able to count the number of solitons included in a given
signal, {\it e.g.,} the one in fig.~\ref{fig:3d}. Indeed, the inset of fig.~\ref{fig:pist} confirms
the presence of two solitons, as expected. Beyond the validity of PIST to {\em detect} KdV solitons in
a periodic system, PIST could be also be applied to directly {\em generate} a KdV soliton gas in
such a geometry.

\section{Conclusion}
We demonstrated the existence of solitons in a system with periodic and curved boundary
conditions. They are observed propagating along a stable torus of fluid (created by a technique we
developed) and are fully characterized (profile, velocity, collision, dissipation and nondispersive
features). These unexplored solitons are found to be governed by a KdV equation with {\em periodic
  boundary conditions} leading to significant differences with infinite straight-line KdV solitons,
such as the observation of subsonic elevation solitons, and the prediction of a nonlinear dependence
of the soliton velocity on its amplitude.  We show that the system periodicity (through the parameter $m$) selects the soliton velocity type (subsonic or supersonic), whereas the Bond number selects the soliton profile (depression or elevation). A nonlinear spectral analysis of solitons is also implemented
(PIST) and is experimentally validated for the first time for a KdV equation with periodic conditions. Our work is not
restricted to hydrodynamics, and thus could be applied to other domains involving periodic systems
governed by a KdV equation. Quantifying the role of dissipation breaking integrability is also of
primary interest~\cite{ChekhovskoyPRL2019}. In the future, this new system could address the
possible existence of KdV soliton gas~\cite{Zakharov1971,ELPRL2005,Costa2014,RedorPRL2019,Suret2020}
in periodic systems, and their collision~\cite{CarboneEPL2016}, as well as of Kaup-Boussinesq
bidirectional solitons~\cite{RedorPRL2019,Zhang2003,Nabelek2020,Congy2021} with corresponding finite-gap
spectral methods~\cite{Smirnov1986}. 



\acknowledgments
 We thank A. Di Palma and Y. Le Goas for technical help on the experimental setup. Part of this
  work was supported by the French National Research Agency (ANR SOGOOD project No. ANR-21-CE30-0061-04), and by a grant from the Simons Foundation MPS No. 651463.


\end{document}